\documentclass{aastex61}
\usepackage{amsmath,amssymb,CJK,lineno,epigraph}
\received{--}
\revised{--}
\accepted{--}
\submitjournal{ApJL}

\shorttitle{\textquoteleft Oumuamua is Hot}
\shortauthors{Ye et al.}

\begin{document}
\begin{CJK*}{UTF8}{gbsn}

\title{1I/2017 U1 (\textquoteleft Oumuamua) is Hot: Imaging, Spectroscopy and Search of Meteor Activity}

\correspondingauthor{Quan-Zhi Ye}
\email{qye@caltech.edu}

\author[0000-0002-4838-7676]{Quan-Zhi Ye (叶泉志)}
\affiliation{Division of Physics, Mathematics and Astronomy, California Institute of Technology, Pasadena, CA 91125, U.S.A.}
\affiliation{Infrared Processing and Analysis Center, California Institute of Technology, Pasadena, CA 91125, U.S.A.}

\author[0000-0002-6702-191X]{Qicheng Zhang}
\affiliation{Division of Geological and Planetary Sciences, California Institute of Technology, Pasadena, CA 91125, U.S.A.}

\author[0000-0002-6702-7676]{Michael S. P. Kelley}
\affiliation{Department of Astronomy, University of Maryland, College Park, MD 20742-2421, U.S.A.}

\author{Peter G. Brown}
\affiliation{Department of Physics and Astronomy, The University of Western Ontario, London, Ontario N6A 3K7, Canada}
\affiliation{Centre for Planetary Science and Exploration, The University of Western Ontario, London, Ontario N6A 5B8, Canada}



\begin{abstract}
1I/2017 U1 (\textquoteleft Oumuamua), a recently discovered asteroid in a hyperbolic orbit, is likely the first macroscopic object of extrasolar origin identified in the solar system. Here, we present imaging and spectroscopic observations of \textquoteleft Oumuamua using the Palomar Hale Telescope as well as a search of meteor activity potentially linked to this object using the Canadian Meteor Orbit Radar. We find that \textquoteleft Oumuamua exhibits a moderate spectral gradient of $10\%\pm6\%~(100~\mathrm{nm})^{-1}$, a value significantly lower than that of outer solar system bodies, indicative of a formation and/or previous residence in a warmer environment. Imaging observation and spectral line analysis show no evidence that \textquoteleft Oumuamua is presently active. Negative meteor observation is as expected, since ejection driven by sublimation of commonly-known cometary species such as CO requires an extreme ejection speed of $\sim40$~m~s$^{-1}$ at $\sim100$~au in order to reach the Earth. No obvious candidate stars are proposed as the point of origin for \textquoteleft Oumuamua. Given a mean free path of $\sim10^9$~ly in the solar neighborhood, \textquoteleft Oumuamua has likely spent a very long time in the interstellar space before encountering the solar system.
\end{abstract}




\section{Introduction}

\epigraph{How delightful it is to have friends coming in from afar!}{\textit{Confucius}}

1I/2017 U1 (\textquoteleft Oumuamua) is likely the first macroscopic object of extrasolar origin identified in the solar system. It was first reported by R. Weryk et al. of the Panoramic Survey Telescope and Rapid Response System (Pan-STARRS) on 2017 Oct 19, and announced with the cometary designation C/2017 U1 (PANSTARRS) based on its orbit \citep{Williams2017a}, but was re-designated as an asteroid under the designation A/2017 U1 due to the lack of cometary activity in deep stacking images taken by several independent observers \citep{Williams2017b,Green2017}. The object was eventually designated as 1I/2017 U1 (\textquoteleft Oumuamua) under a new designation system proposed for interstellar objects \citep{Williams2017d}. As of 2017 Nov 6, the International Astronomical Union's Minor Planet Center gives a hyperbolic orbital solution with eccentricity $e=1.197$ and hyperbolic excess speed $v_\infty=26$~km~s$^{-1}$ \citep{Williams2017c}.

\textquoteleft Oumuamua's visit provides an unprecedented opportunity to directly study an extrasolar planetesimal at close range. \textquoteleft Oumuamua passed relatively closely to the Earth, with a minimum distance of 0.161~au and minimal orbit intersection distance (MOID) of 0.096~au. This trajectory not only aided observation for Earth-based observers, but also permits potential dust ejected by the object (if any) to reach the Earth and appear as meteors. Here, we present our telescopic and meteor observations of \textquoteleft Oumuamua.

\section{Imaging}

We obtained direct imagery of \textquoteleft Oumuamua with the Large Format Camera (LFC) on the Palomar 5-m Hale Telescope, on 2017 Oct 26 02:12--02:21 UT (Table~\ref{tbl:obs}). The LFC camera is a mosaic of six 2k$\times$4k CCDs located at the prime focus of the Hale Telescope. It has a field diameter of 24' and a 2$\times$2 binned pixel scale of 0.36''~pixel$^{-1}$. We obtained $2\times90$~s on-target exposures in each of $r'$ and $g'$, all calibrated with bias and flat frames taken earlier the same night. The object showed no significant deviation from a point source in any individual frames.

\begin{figure}
\includegraphics[width=0.5\textwidth]{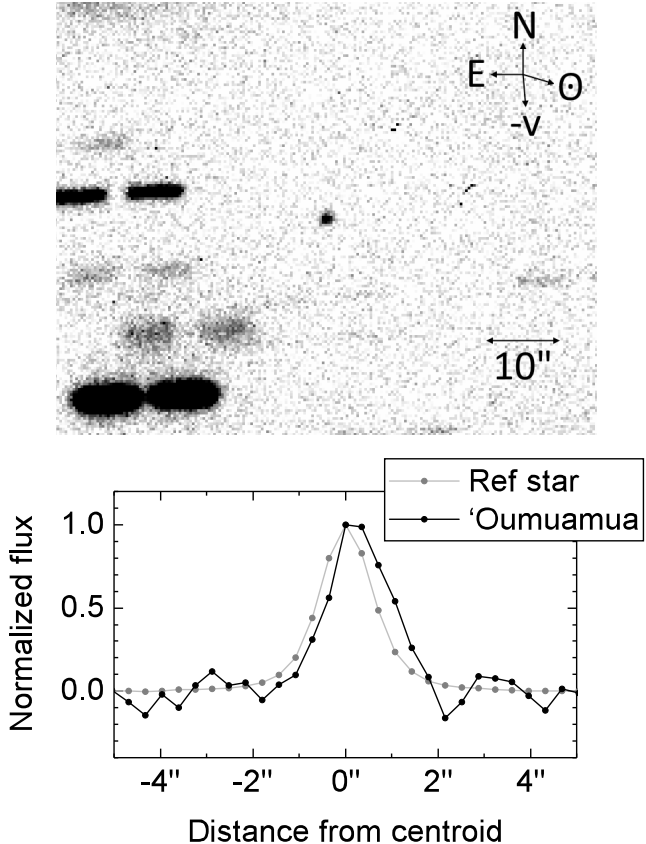}
\caption{Upper panel: composite $g'$+$r'$ image of \textquoteleft Oumuamua. Directions to north, east, minus heliocentric velocity vector and the direction to the Sun are marked by arrows. Lower panel: surface brightness profile of \textquoteleft Oumuamua (black curve) and a nearby reference star (grey curve). The peak of the profile of \textquoteleft Oumuamua is slightly broadened, causing a 0.5-pixel offset of the profile to the right, but the width of the profile is comparable to the reference star.}
\label{fig:img}
\end{figure}

We then proceeded to search for cometary activity exhibited by \textquoteleft Oumuamua. Our total integration time (6~min.) is admittedly quite short compared to that of typical searches (usually several tens of minutes), but as we will show below, the relatively large aperture size of the Hale Telescope permits useful result to be derived. To increase the signal-to-noise ratio, we combined all $r'$ and $g'$ frames into a composite image. One-dimensional surface brightness profiles of \textquoteleft Oumuamua and a nearby reference star were then obtained by averaging the pixels along the direction of the object's motion and subtracting the sky background. As shown in Figure~\ref{fig:img}, \textquoteleft Oumuamua appeared completely stellar, a result consistent with its present designation. We performed aperture photometry on the object using SDSS DR12 \citep{Alam2015} as reference. We found AB magnitudes of $r'=21.47\pm0.06$ and $g'=22.07\pm0.22$, giving $g'-r'=0.60\pm0.23$, which is grossly consistent with a reddish color. The $1\sigma$ bound on excess surface brightness in the 3.6--5.0'' annulus is 28.1~mag~arcsec$^{-1}$, corresponding to an $Af\rho$ upper limit of $\sim2\times10^{-4}$~m. Here $Af\rho$ \citep{Ahearn1984} is a proxy of dust production rate of comets. Typical $Af\rho$ values for comets vary from $10^{-2}$ to 100~m \citep{Ahearn1995}. The AB magnitudes we reported are likely undermined by the rotation of \textquoteleft Oumuamua. Our observation is likely too short to cover the entire rotation. Independent time-series photometry made by \citet{Knight2017} shows a relatively long rotation period ($\gtrsim5$~hr) and a moderate light-curve amplitude ($\gtrsim1$ mag), which is not atypical for solar system asteroids of similar sizes.

\begin{table*}
\begin{center}
\caption{Circumstances of the imaging and spectroscopic observations.\label{tbl:obs}}
\begin{tabular}{lcccccc}
\hline
Date (UT) & Instrument & $r_\mathrm{H}$\tablenotemark{a} & $\varDelta$\tablenotemark{b} & $\alpha$\tablenotemark{c} & Airmass & Zenith seeing \\
 & & (au) & (au) & & \\
\hline
2017 Oct 26 02:12--02:21 & Hale + LFC & 1.386 & 0.432 & $20.8^\circ$ & 1.76--1.65 & 1.3'' \\
2017 Oct 26 04:47--05:30 & Hale + DBSP & 1.389 & 0.436 & $20.9^\circ$ & 1.16--1.14 & 1.5'' \\
\hline
\end{tabular}
\end{center}
\tablenotetext{a}{Heliocentric distance.}
\tablenotetext{b}{Geocentric distance.}
\tablenotetext{c}{Phase angle.}
\end{table*}

\section{Spectroscopy}

We obtained an optical spectrum of \textquoteleft Oumuamua using the Double Spectrograph (DBSP) on the Hale Telescope on 2017 Oct 26 from 04:47--05:30 UT (Table~\ref{tbl:obs}). The DBSP consists of two arms split by a dichroic module into blue and red channels. We used a 600~lines~mm$^{-1}$ grating for the blue channel and 316~lines~mm$^{-1}$ grating for the red channel, which provided a spectral resolving power of 917 and 912 at blaze angles of 378 and 715~nm, respectively. A slit of 1.5'' was used in accordance to the seeing condition at the time of the observation, and was aligned along the parallactic angle in order to reduce the effect of atmospheric differential refraction. We obtained $4\times10$~min of useful on-target integration time. A flux standard star (BD+28~4211) and a solar analog star (HD~1368) were also observed around the same time at similar airmass conditions, in order to allow atmospheric correction and to derive a reflectance spectrum, respectively.

The data are calibrated using the bias and flat fields taken earlier in the night. Wavelength calibrations are then performed using a Fe-Ar lamp for the blue channel and a He-Ne-Ar lamp for the red channel. After calibration, individual exposures of \textquoteleft Oumuamua are average-combined in order to get the final spectrum. We then extract spectra of the target, solar analog and flux standard star using a 1.5''-wide aperture, and perform flux calibration for both the target and the solar analog using the flux standard star. The final reflectance spectrum is obtained by dividing the flux-calibrated target spectrum by the solar analog spectrum.

\begin{figure*}
\includegraphics[width=\textwidth]{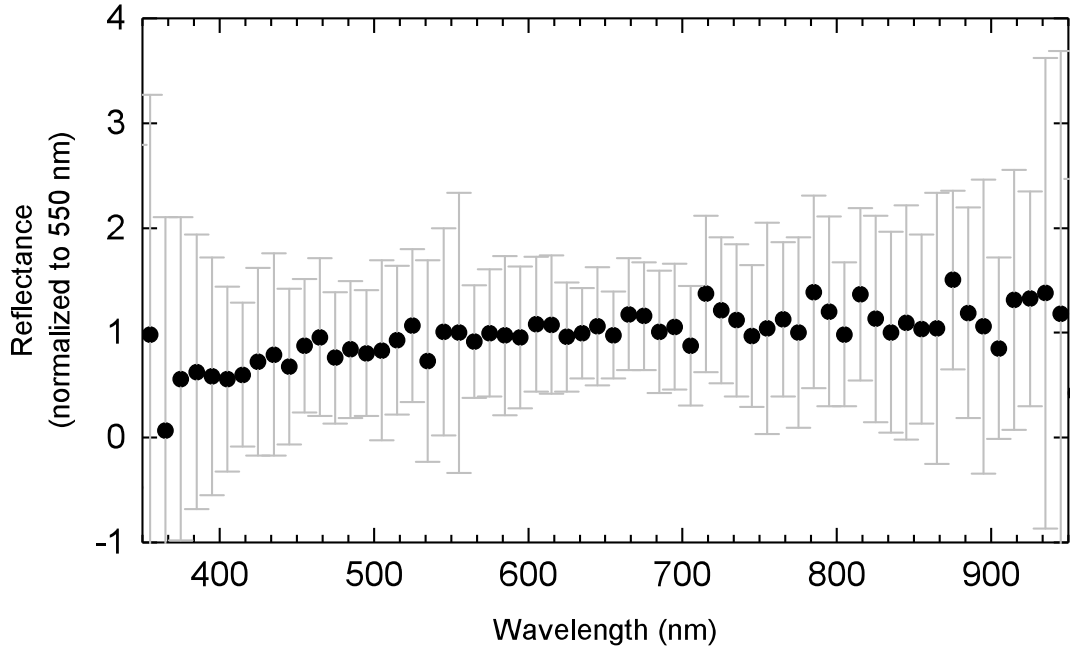}
\caption{Reflectance spectrum of \textquoteleft Oumuamua binned to 10~nm and weighted by the uncertainty of each bin.}
\label{fig:spec}
\end{figure*}

The calibrated spectrum of \textquoteleft Oumuamua is shown in Figure~\ref{fig:spec}. The spectrum is normalized to the reflectance at 550~nm. Our result is in general agreement with the spectrum taken by \citet{Masiero2017} a day earlier, with a reddish, featureless spectrum across the entire wavelength window, though our spectrum has a shorter wavelength cutoff due to improved atmospheric conditions (375 nm versus 520 nm).

Spectral gradient is a useful metric to understand the surface composition of a small body. The normalized reflectance gradient can be calculated by $S'=(\mathrm{d}S/\mathrm{d}\lambda)/\bar{\mathbf{S}}$, where $S$ is the reflectance, and $\bar{\mathbf{S}}$ is the mean reflectance in the wavelength range used in the calculation \citep{Jewitt2002}. From the reflectance spectrum presented above, we derive $S'=10\%\pm6\%~(100~\mathrm{nm})^{-1}$ at 650~nm considering the wavelength range 400--900~nm. This value is in line with the broadband color derived above and with the gradient reported by \citet{Fitzsimmons2017}. We note that this gradient encompasses the classes of dead and active cometary nuclei, Trojans and active Centaurs, but is noticeably less red than in active Centaurs and all classes of Kuiper-belt objects (KBOs), which have average $S'=23\%\pm2\%~(100~\mathrm{nm})^{-1}$ \citep{Jewitt2015}.

The flux-calibrated spectrum also allows us to constrain the emission intensity of major cometary species observed in the optical: CN, C$_2$ and C$_3$, centered at 387, 514 and 406~nm respectively. In accordance to the width of the emission lines, we measure the flux of CN and C$_3$ using a 5~nm aperture and C$_2$ using a 10~nm aperture, which yields $3\sigma$ upper limits of $7.0\times10^{-20}$, $8.4\times10^{-20}$, and $4.4\times10^{-20}~\mathrm{W~m^{-2}}$ for CN, C$_2$ and C$_3$, respectively. Using the technique described by \citet{Farnham2000}, these numbers can be translated into $Q$(CN)$<2\times10^{22}~\mathrm{molecule~s^{-1}}$, $Q$(C$_2$)$<4\times10^{22}~\mathrm{molecule~s^{-1}}$, and $Q$(C$_3$)$<2\times10^{21}~\mathrm{molecule~s^{-1}}$ in terms of production rates. These upper-limits are comparable to the activity level of some of the most weakly active comets ever measured, such 209P/LINEAR \citep{Schleicher2016}.

\section{Search for meteor activity}

\textquoteleft Oumuamua's orbit has a MOID of $\sim0.1$~au with the Earth, a distance less than the MOID of parent bodies with known meteor showers visible at the Earth \citep[e.g.,][]{Drummond1981}. If \textquoteleft Oumuamua has an accompanying dust/meteoroid stream of sufficient spatial density and extending 0.1~au from its orbit at its node, some meteor activity might be visible at the Earth, particularly as \textquoteleft Oumuamua passed its nodal point near the time of Earth's closest approach to the object's orbit.

We use the approach of \citet{Neslusan1998} to calculate the theoretical radiant and timing of potential meteor activity from \textquoteleft Oumuamua. Meteor activity, if any, would occur near 2017 Oct 18.0 UT (solar elongation $204.6^\circ$), from a geocentric radiant of $\alpha=160^\circ$, $\delta=-8^\circ$ with geocentric speed $v_\mathrm{G}=65$~km~s$^{-1}$ (J2000). This radiant is in the constellation of Sextans, which, on Oct 18, was at a solar elongation of $\sim50^{\circ}$, making the radiant only briefly visible in dark skies towards sunrise at northern latitudes.

We examine the data collected by the Canadian Meteor Orbit Radar (CMOR), an interferometric radar array located near London, Canada. The details of CMOR operations and analysis can be found in \citet{Jones2005, Brown2008a} and \citet{Weryk2012}. The search for shower activity from the theoretical radiant for \textquoteleft Oumuamua was performed using the 3-dimensional wavelet analysis \citep[c.f.][]{Bruzzone2014}, which is useful for the detection of weak meteor activities that are not easily identifiable from conventional radiant plots \citep[e.g.][]{Sato2017}. We note that the radiant as seen from CMOR was only above the horizon from 8--19 UT on Oct 18; no meteors from the time of closest approach near 2 UT were detectable from CMOR. 

\begin{figure}
\includegraphics[width=0.5\textwidth]{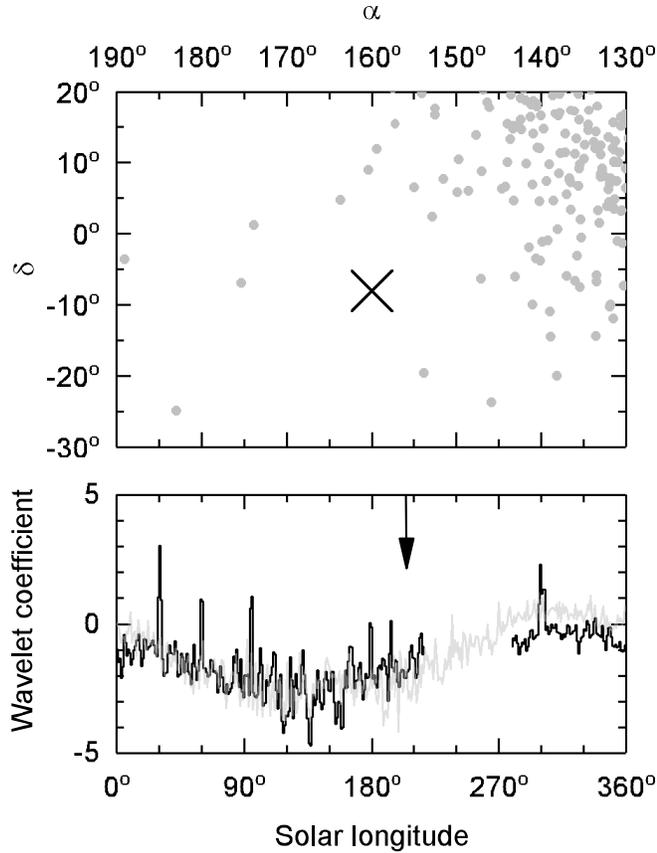}
\caption{Upper panel: distribution of geocentric meteor radiants within $\pm10\%$ of the predicted speed ($v_\mathrm{G}=65$~km~s$^{-1}$) within the interval 2017 Oct 17--19 UT around the location of the predicted theoretical radiant for any meteor activity which might be associated with \textquoteleft Oumuamua. This theoretical radiant is marked with a cross. Lower panel: change of the wavelet coefficient (a proxy of meteoroid flux in arbitrary units) from 2017 Jan 1 to Oct 31 (black curve) and 2002--2016 (grey curve), with the predicted timing of the meteor activity originated from \textquoteleft Oumuamua marked by an arrow.}
\label{fig:met}
\end{figure}

Figure~\ref{fig:met} shows the distribution of meteor radiants within $\pm10\%$ of the predicted speed and within $\pm1$~day of the predicted timing for any shower produced by \textquoteleft Oumuamua, as well as the variation of the wavelet coefficient (a proxy of the meteoroid flux) at the predicted radiant throughout the year. We do not see any significant enhancement at the predicted timing and radiant; indeed, this level of activity is within the noise floor at this radiant location for CMOR data collected between 2002--2016. 

Accounting for CMOR's collecting area and detection sensitivity at an arrival speed 65~km~s$^{-1}$ following the procedure in \citet{Ye2016}, we estimate that the detection limit of the meteoroid flux at this radiant position is $\sim10^{-3}~\mathrm{km^{-2} \cdot hr^{-1}}$, appropriate to a limiting mass of $\sim10^{-8}$~kg \citep{Ye2016b}. If we assume isotropic ejection from \textquoteleft Oumuamua, this translates to a limit of the dust production rate of $\lesssim10$~kg~s$^{-1}$ at the source. This upper limit is within the range of typical cometary dust production rates, which varies from a few kg~s$^{-1}$ to $\sim10^5$~kg~s$^{-1}$.

To understand the age of the potentially observable meteors, we simulate the dynamical evolution of radar-sized dust ($\sim100~\micron$-sized) ejected at different heliocentric distances and speeds, using the dust dynamical code developed in our earlier work \citep{Ye2016c}. The code accounts for gravitational perturbation by major planets (from Mercury through Neptune, with the Earth-Moon system represented by a single perturber) as well as radiation pressure from the Sun. We find that dust ejected at a modest speed of 1~m~s$^{-1}$ need $\sim600$~yr to reach an Earth-intercepting trajectory. As \textquoteleft Oumuamua was about 3000~au from the Sun 600~yr before its perihelion, only H$_2$ ice could have started sublimation \citep{Meech2004}, a process that is yet to be directly observed in the solar system. Ejections driven by sublimation of commonly-known cometary species such as CO (onset distance $\sim120$~au) need to have ejection speed of $\sim40$~m~s$^{-1}$ in order to reach the Earth. Currently there is no known mechanism that can power such energetic ejection at such large heliocentric distance. Thus the absence of meteor activity is as expected for \textquoteleft Oumuamua. We also note that meteor observation is only sensitive to dust produced well before the perihelion passage, as dust produced near the Sun would not have enough time to reach the Earth, given typical ejection speeds for radar-sized meteoroids at $\sim1$~au ($<100$~m~s$^{-1}$). 

\section{Discussion}

The moderate spectral gradient of \textquoteleft Oumuamua indicates that its surface is devoid of ultrared material that is common on outer solar system objects like KBOs. Various classes of KBOs are typically very red in color, with a spectral gradient $\gtrsim20\%~(100~\mathrm{nm})^{-1}$, likely due to the irradiation of organic material resulting from the bombardment of energetic particles \citep{Brunetto2006}. The less reddish color of \textquoteleft Oumuamua suggests that the object was either formed close to its original central star, or has lost its ultrared material due to close approach(es) to its original or other stars. It is difficult to say which scenario is more likely due to the chaotic nature of small body dynamics. For the case of the solar system, it is known that planetary perturbations occasionally send small bodies out of the solar system. Known examples include D/1770 L1 (Lexell) and C/1980 E1 (Bowell) \citep{Lexell1779,Bowell1980a}.

Can we trace the origin of \textquoteleft Oumuamua? \citet{Mamajek2017} has shown that the velocity and trajectory of \textquoteleft Oumuamua is consistent with a typical interstellar object (ISO) drawn from the velocity distribution of the local stellar population, but noted no definite star of origin. We conducted a scan of stellar close approaches to the nominal trajectory with the Gliese star catalog \citep{Gliese1991}, which also reveals no obvious candidates in the immediate vicinity of the solar system. Close encounters of an ISO to multiple planetary systems is extremely rare, considering that the mean free path of an ISO in the solar neighborhood is $l=(\pi R^2 \rho)^{-1}\approx10^9$~ly, assuming $R=10$~au for the encounter distance to a planetary system \citep[chosen in accordance to the distance that ultrared objects start to disappear in the solar system, e.g.,][though there can be a factor of 10 difference depending on the type of the host star]{Melita2012}, and the stellar density $\rho=0.004~\mathrm{ly^{-3}}$ for the solar neighborhood. This translates to a travel time of $10^{13}$~yr at a speed comparable to its relative motion through the solar neighborhood. Presently, the positional error grows to the average stellar distance in $\sim10^7$~yr. All these factors make it difficult on pinpointing the point of origin for \textquoteleft Oumuamua, though the large mean free path also imply that the solar system is likely the first planetary system that \textquoteleft Oumuamua encountered besides its birth planetary system. If a past close encounter to a planetary system can be found, that system is likely the true point of origin for \textquoteleft Oumuamua.

\section{Conclusion}

We presented imaging and spectroscopic observations of the potential interstellar object 1I/2017 U1 (\textquoteleft Oumuamua). The object appeared completely stellar in our images, with $Af\rho<2\times10^{-4}$~m, consistent with its current designation. The optical spectrum revealed a moderate spectral gradient of $10\%\pm6\%~(100~\mathrm{nm})^{-1}$, consistent with a small body residing in a warmer environment susceptible to the depletion of organic material. Plausible explanations include a formation in the inner region of a protoplanetary disk, or previous close encounters with stars. From the spectrum, we determined upper limits to the production rates of CN, C$_2$ and C$_3$: $Q(\text{CN})<2\times10^{22}~\mathrm{molecule~s^{-1}}$, $Q(\text{C}_2)<4\times10^{22}~\mathrm{molecule~s^{-1}}$, and $Q(\text{C}_3)<2\times10^{21}~\mathrm{molecule~s^{-1}}$. These limits are comparable to the activity level of weakly active comets in the known comet population.

We also searched radar meteor data for meteor activity that could have originated from a recent ejection from \textquoteleft Oumuamua, without any positive detection. By applying a dust dynamical model and assuming an ejection speed comparable to gravitational escaping speed, we concluded that the dust production rate of \textquoteleft Oumuamua is $\lesssim10$~kg~s$^{-1}$ at a solar distance of $\sim10^3$~au. Ejection driven by sublimation of commonly-known cometary species such as CO requires an extreme ejection speed of $\sim40$~m~s$^{-1}$ at $\sim100$~au in order to reach the Earth.

The prospects for tracing the point of origin of \textquoteleft Oumuamua are slim. Despite the efforts made by the authors and others, no obvious candidates have been proposed in the immediate vicinity of the solar system. Given the stellar density in the solar neighborhood, interstellar objects like \textquoteleft Oumuamua can travel $10^9$~ly before having a close encounter with a planetary system. Our knowledge of the inbound trajectory of \textquoteleft Oumuamua is also hampered by the fact that the object was only discovered in its outbound phase.

The discovery of what is likely the first macroscopic interstellar object is nevertheless encouraging. Next generation time-domain sky surveys, such as the Large Synoptic Sky Survey \citep[LSST;][]{Tyson2002}, will provide deeper coverage over wider areas, hopefully revealing more objects like \textquoteleft Oumuamua \citep{Cook2016,Engelhardt2017}. This will provide better estimates of the number density and size distribution of interstellar objects which are presently poorly constrained.

\acknowledgments

We thank an anonymous reviewer for his/her comments, Nadia Blagorodnova for kindly sharing her Palomar observation time with us and helping with the observation, Joe Masiero for helping us understand the operation of the Hale Telescope, as well as Kajsa Peffer and Paul Nied for observational support. Q.-Z. is supported by the GROWTH project (National Science Foundation Grant No. 1545949). This research makes use of observations from the Hale Telescope at Palomar Observatory, which is owned and operated by Caltech and administered by Caltech Optical Observatories, as well as data and services provided by the International Astronomical Union's Minor Planet Center. Funding support from the NASA Meteoroid Environment Office (cooperative agreement NNX15AC94A) for CMOR operations is gratefully acknowledged.

%

\vspace{5mm}
\facilities{CMOR, Hale (LFC and DBSP)}


\software{Astropy \citep{Astropy2013},  
          IRAF \citep{Tody1986}
          }

\end{CJK*}



\end{document}